# An $O(n)$ Time Algorithm For Maximum Induced Matching In Bipartite $Star_{123}$-free Graphs

Ruzayn Quaddoura

Department of Computer Science, Faculty of Science and Information Technology
Zarqa University
Zarqa-Jordan

Abstract— A matching in a graph is a set of edges no two of which share a common vertex. A matching $M$ is an induced matching if no edge connects two edges of $M$. The problem of finding a maximum induced matching is known to be NP-hard in general and specifically for bipartite graphs. Lozin has been proposed an $O(n^3)$ time algorithm for this problem on the class of bipartite $Star_{123}, Sun_4$-free graphs. In this paper we improve and generalize this result in presenting a simple $O(n)$ time algorithm for maximum induced matching problem in bipartite $Star_{123}$-free graphs.

Keywords-Bipartite graph; Decomposition of graphs; Design and analysis of algorithms; Matching; Induced Matching.

## I. INTRODUCTION

A *matching M* of a graph $G = (V, E)$ is a subset of edges with the property that no two edges of $M$ share a common vertex. A matching is called *induced* if no two edges in the matching have a third edge connecting them. Equivalently, the subgraph of $G$ induced by $M$ consists of exactly $M$ itself. We study the problem of finding in $G$ an induced matching of maximum cardinality, denoted $i\mu(G)$. This problem has been introduced by Cameron [3], where he has proved its NP-hardness in the class of bipartite graphs. The maximum induced matching problem was shown to be polynomial for several classes of graphs: for chordal graphs and for interval graphs by Cameron [3], for circular-arc graphs by Golumbic and Laskar [5], and for trapezoid graphs, $k$-interval-dimension graphs and cocomparability graphs by Golombic and Lewenstein [4]. Fricke and Laskar give a linear algorithm for trees [1]. Lozin in [8] describes an $O(n^3)$ time algorithm for the problem on bipartite $Star_{123}, Sun_4$-free graphs where $n$ is the number of vertices. In addition, he studied in [7] the class of bipartite $Star_{123}$-free graphs and conjectured that his result in [8] can be extended to this class of bipartite graphs. In this paper we improve and generalize Lozin's algorithm in presenting a simple $O(n)$ time algorithm for this problem on bipartite $Star_{123}$-free graphs. Our algorithm is based on the recognition algorithm of the class $Star_{123}$-free bipartite graphs introduced by Quaddoura in [6].

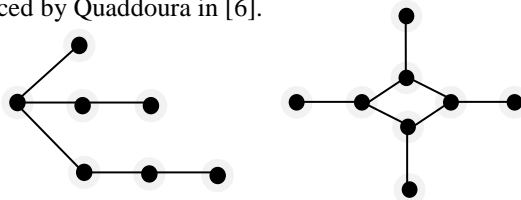

$Star_{123}$         $Sun_4$
Figure. 1. **$Star_{123}$** and **$Sun_4$** configurations

## II. DEFINITION AND PROPERTIES

A bipartite graph $G = (B \cup W, E)$ is defined by two disjoint vertex subsets $B$ – the black vertices and $W$ - the whites ones, and a set of edges $E \subseteq B \times W$. The *bi-complement* of a bipartite graph $G = (B \cup W, E)$ is the bipartite graph defined by $\bar{G}^{bip} = (B \cup W, B \times W - E)$. If the color classes $B$ and $W$ are both non empty the graph will be called *bichromatic*, *monochromatic* otherwise. A vertex $x$ will be called *isolated* (resp. *universal*) if $x$ has no neighbors in $G$ (resp. in $\bar{G}^{bip}$). A *complete* bipartite graph is a graph having only universal white vertices and universal black vertices. A *stable* set is a set of isolated vertices. A chordless path on $k$ vertices is denoted by $P_k$ and a chordless cycle on $k$ vertices is denoted by $C_k$. Given a subset $X$ of the vertex set $V(G)$, the subgraph induced by $X$ will be denoted by $G[X]$ or simply by $X$ if there is no confusion. A $K_2$ is a complete bipartite graph with two vertices. A $2K_2$ is a two copies of a $K_2$.

**Definition 1** [2] *Given a bipartite graph $G = (B \cup W, E)$ of order at least 2, G is $K + S$ graph if and only if G contains an isolated vertex or its vertex set can be decomposed into two sets K and S such that K induces a complete bipartite graph while S is a stable set.*

**Property 1** [2] *Let $G = (B \cup W, E)$ be a bipartite graph of order at least 2. G is $K + S$ graph if and only if there exists a partition of its vertex set into two non empty classes $V_1$ and $V_2$ such that all possible edges exists between the black vertices of $V_1$ and the white vertices of $V_2$ while there is no edge connecting a white vertex of $V_1$ with a black vertex of $V_2$.*

Such partition is referred as associated partition of $G$ and is denoted by the ordered pair $(V_1, V_2)$.

**Property 2** [2] *A bipartite graph $G$ is a $K + S$ graph if and only if $G$ admit a unique (up to isomorphism) partition of its vertex set $(V_1 \cup V_2 ... \cup V_k)$ satisfying the following conditions:*

    *a)* $\forall i = 1, ..., k-1, (V_1 \cup ... \cup V_i, V_{i+1} \cup ... \cup V_k)$ *is an associated partition to the graph $G$*

    *b)* $\forall i = 1, ..., k, G[V_i]$ *is not a $K + S$ graph.*

The partition $(V_1, ..., V_k)$ of the above property is called $K + S$ decomposition while a set $V_i$ said to be $K + S$ -component of the graph.

From $K + S$ decomposition together with the decomposition of bipartite graph $G$ into its connected components (parallel decomposition) or those of $\bar{G}^{bip}$ (series decomposition) yield a new decomposition scheme for $G$ called *canonical decomposition*. It is show in [2] that whatever the order in which the decomposition operators are applied ($K + S$ decomposition, series decomposition or parallel decomposition), a unique set of indecomposable graphs with respect to canonical decomposition is obtained. Obviously, a unique tree is associated to this decomposition. The internal nodes are labeled according to the type of decomposition applied, while every leaf correspond to a vertex of $G$. Hence there are four types of internal nodes, parallel node (labeled $P$), series node (labeled $S$), $K + S$ node (labeled $K + S$), and indecomposable node (labeled $N$). By convention, the set of vertices corresponding to the set of leaves having an internal node $\alpha$ as their least common ancestor as well as the subgraph induced by this set of leaves will be denoted simply by $\alpha$.

**Observation 1** let $G$ be a bipartite graph and $T$ be its canonical decomposition tree. According to the order in which the decomposition operations are applied, every child of a $P$-node or a $S$-node cannot be a vertex. Such node would have either an isolated or a universal vertex and thus would induce a $K + S$ graph.

Following the recognition algorithm given in [6], bipartite $Star_{123}$ -free graphs are bipartite graphs whose indecomposable graphs within canonical decomposition are reduced to signal vertices or to an extended path $EP_k$ or the bi-complement of an extended path $EP_k$ or an extended cycle $EC_k$ or the bi-complement of an extended cycle $EC_k$. In all cases $k \geq 7$. More precisely

**Definition 2** [6] *A graph $G$ is said to be an extended path $EP_k$ if there is a partition of the vertex set of $G$ into a monochromatic sets $\{V_1, ..., V_k\}$ such that $E = \bigcup_{i=1}^{k-1} V_i \times V_{i+1}$ and $k \geq 7$.*

**Definition 3** [6] *A graph $G$ is said to be an extended cycle $EC_k$ if there is a partition of the vertex set of $G$ into a*

monochromatic sets $\{V_1, ..., V_k\}$ such that $E = \bigcup_{i=1}^{k-1} V_i \times V_{i+1} \cup V_1 \times V_k$ and $k \geq 7$.

The construction of the canonical decomposition tree of a bipartite $Star_{123}$-free graph tree can be obtained in linear time from the algorithm given by Quaddoura in [6]. According to this algorithm, every child of a $N$-node is a node marked by $P'$ corresponding to a set $V_i, i = 1 ... k$, if $|V_i| > 1$, or to a vertex of $G$ otherwise. Figure 2 illustrate a bipartite $Star_{123}$-free graph and its canonical decomposition tree.

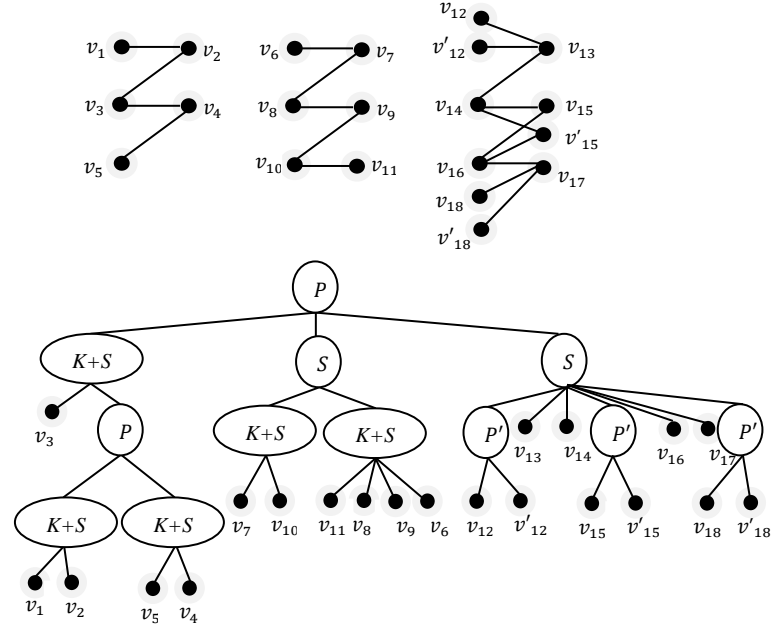

Figure. 2. A bipartite $\textbf{\textit{Star}}_{123}$-free graph and its canonical decomposition tree

### III. MAXIMUM INDUCED MATCHING IN BIPARTITE $Star_{123}$-FREE GRAPHS

Let $G$ be a bipartite $Star_{123}$-free graph and $T(G)$ be its canonical decomposition tree. Our algorithm uses post order traversal to visit all the nodes of $T(G)$. Whenever an internal node $\alpha$ is visited, we compute a maximum induced matching of the subgraph induced by $\alpha$ from the maximum induced matching's of its children say $\alpha_1, \alpha_2, ..., \alpha_k$. For this purpose we distinguish several cases according to the type of $\alpha$.

Obviously, If $\alpha$ is a $P$-node then $i\mu(\alpha) = \cup_{i=1}^{k} i\mu(\alpha_i)$. Also, if $\alpha$ is a $P'$-node then $i\mu(\alpha) = \emptyset$.

A set of vertices $A$ is called *module* if every vertex in $A$ has the same neighborhood outside of $A$. A bipartite graph whose every module is of size 1 will be called *prime*. It is not hard to see that any bipartite graph $G$ has a unique (up to isomorphism) maximal prime induced subgraph that can be obtained by choosing exactly one vertex in each module of $G$. Lozin in [8] proved the following Lemma.

**Lemma 1** *If $H$ is a maximal prime induced subgraph of a graph $G$, then $i\mu(G) = i\mu(H)$.*

Suppose now $\alpha$ is a $N$-node. As mentioned above, $\alpha$ induces an extended path $EP_k$ or its bi-complement or an extended cycle $EC_k$ or its bi-complement. Clearly, in this case, the maximal prime induced subgraph of $\alpha$ is a path $P_k$ or its bi-complement (if $\alpha$ induces an extended path $EP_k$ or its bi-complement) or a cycle $C_k$ or its bi-complement (if $\alpha$ induces an extended cycle $EC_k$ or its bi-complement). The following simple Lemma is proved in [8].

**Lemma 2** $|i\mu(P_k)| = \lfloor(k+1)/3\rfloor$, $|i\mu(C_k)| = \lfloor k/3\rfloor$. Let $k \geq 7$ then $\left|\mu(\bar{P}_k^{bip})\right| = \left|i\mu(\bar{C}_k^{bip})\right| = 2$.

By Lemma 2, the set $\{v_{3i-2}v_{3i-1}: 1 \leq i \leq \lfloor(k+1)/3\rfloor\}$ is a maximum induced matching of the path $P_k = v_1v_2 \dots v_k$, the set $\{v_{3i-2}v_{3i-1}: 1 \leq i \leq \lfloor k/3\rfloor\}$ is a maximum induced matching of the cycle $C_k = v_1v_2 \dots v_k$, and the set $\{v_1v_4, v_2v_5\}$ is a maximum induced matching of $\bar{P}_k^{bip}$ or $\bar{C}_k^{bip}$.

Let's discus the cases when $\alpha$ is a $S$-node or a $K + S$-node.

**Lemma 3** *Let $\alpha$ be a $K + S$ node. Then*
*a) If every child of $\alpha$ is a vertex then $|i\mu(\alpha)| \leq 1$*
*b) Else $i\mu(\alpha) = i\mu(\alpha_j)$ where $\alpha_j$ is the child of $\alpha$ which satisfies $|i\mu(\alpha_j)| = max\{|i\mu(\alpha_i)|: \alpha_i \text{ is not a vertex}\}$*

*Proof* By Observation 1 the father of a leaf is either a $N$-node, a $P'$-node or a $K + S$-node. The validity of Lemma deduces directly from Property 3 by remarking that there is no $2K_2$ or $\alpha$ can share vertices with two different children. □

**Lemma 4** *Assume that $\alpha$ is a $S$-node.*
*a) If any child $\alpha_i$ of $\alpha$ satisfies that $|i\mu(\alpha_i)| \leq 1$ then $|i\mu(\alpha)| = 2$*
*b) Else $i\mu(\alpha) = i\mu(\alpha_j)$ where $\alpha_j$ is the child of $\alpha$ which satisfies $|i\mu(\alpha_j)| = max\{|i\mu(\alpha_i)|: 1 \leq i \leq k\}$.*

*Proof* The Lemma can be deduced from the following Claim :

*Claim* the cardinality of maximum induced matching of $\alpha$ which shares vertices between different children is 2

*Proof* Let $X$ denotes to such induced matching. Every child of $\alpha$ contains two nonadjacent vertices of different color, otherwise $\alpha$ would contain a universal vertex and hence, by Observation 1, $\alpha$ is a $K + S$-node, a contradiction. Let $v_1, v_2$ be two nonadjacent vertices of a child say $\alpha_1$ of $\alpha$ such that $v_1, v_2$ are of different color, and $v_3, v_4$ be two non adjacent vertices of a child say $\alpha_2$ distinct of $\alpha_1$ and $v_3, v_4$ are of different color, then the set $\{v_1, v_2, v_3, v_4\}$ induces a $2K_2$.

Therefore, $X \geq 2$. Since $\alpha$ is a $S$-node, any vertex $v$ of a child distinct of $\alpha_1$ and $\alpha_2$ is adjacent to the two vertices of $\{v_1, v_2, v_3, v_4\}$ whose have the same color, so $X = 2$.

The proof of Lemma 4 shows that any child of a $S$-node contains two nonadjacent vertices of different color. The following Lemma allow us to find easily these two vertices when any child $\alpha_i$ of $\alpha$ satisfies that $|i\mu(\alpha_i)| \leq 1$.

**Lemma 4** *if $|i\mu(\alpha)| \leq 1$ then one of the following is hold:*
*a) $\alpha$ is a vertex.*
*b) $\alpha$ is a $P'$ node.*
*c) $\alpha$ is a $K + S$-node and every child of $\alpha$ is a vertex.*

*Proof* If $\alpha$ is a $N$-node or a $S$-node then by Lemma 2 and Lemma 4, $|i\mu(\alpha)| \geq 2$. So it is enough to prove that $\alpha$ cannot be a $P$-node. Suppose that every child of $\alpha$ contains at least one edge, otherwise $\alpha$ would contain an isolated vertex and hence $\alpha$ would be a $K + S$-node. Therefore $|i\mu(\alpha)| \geq 2$, a contradiction. □

The above discussion leads us to the following algorithm

---

*Algorithm Maximum Induced Matching*
Input : A bipartite $Star_{123}$-free graph $G$ and its canonical decomposition tree $T(G)$.
Output : A maximum induced matching $i\mu(G)$.

Let $\alpha$ be a node on a post order traversal of $T(G)$
if $\alpha$ is a vertex or a $P'$-node then $i\mu(G) = \emptyset$
else let $\alpha_1, \alpha_2, \dots, \alpha_k$ be the children of $\alpha$
  if $\alpha$ is a $N$-node then for every child $\alpha_i$ of $\alpha$, pick a vertex $v_i, 1 \leq i \leq k$
    if $\alpha$ induces an $EP_k$ then $i\mu(G) = \{v_{3i-2}v_{3i-1}: 1 \leq i \leq \lfloor(k+1)/3\rfloor\}$
    if $\alpha$ induces an $EC_k$ then $i\mu(G) = \{v_{3i-2}v_{3i-1}: 1 \leq i \leq \lfloor k/3\rfloor\}$
    if $\alpha$ induces an $\overline{EP}_k^{bip}$ or an $\overline{EC}_k^{bip}$ then $i\mu(G) = \{v_1v_4, v_2v_5\}$
  else if $\alpha$ is a $K + S$-node then
    if every child of $\alpha$ is a vertex then
      if there is two adjacent vertices $v_1, v_2$ of $\alpha$ then $i\mu(G) = \{v_1v_2\}$
      else $i\mu(G) = \emptyset$
      if $\alpha$ contains two nonadjacent vertices $v_1, v_2$ of different color then $I_\alpha = \{v_1, v_2\}$
    else $i\mu(G) = i\mu(\alpha_j)$ where $\alpha_j$ is the child of $\alpha$ satisfying $|i\mu(\alpha_j)| = max\{|i\mu(\alpha_i)|: \alpha_i \text{ is not a vertex}\}$
  else if $\alpha$ is a $S$-node then
    if for every $1 \leq i \leq k, |i\mu(\alpha_i)| \leq 1$ then
      let $I_{\alpha_1} = \{v_1, v_2\}$ and $I_{\alpha_2} = \{v_3, v_4\}$ such that $v_1, v_3$ are black vertices and $v_2, v_4$ are white, $i\mu(G) = \{v_1v_4, v_2v_3\}$
    else $i\mu(G) = i\mu(\alpha_j)$ where $\alpha_j$ is the child of $\alpha$ satisfying $|i\mu(\alpha_j)| = max\{|i\mu(\alpha_i)|: 1 \leq i \leq k\}$

else //$\alpha$ is a *P*-node// $i\mu(G) = \cup_{i=1}^{k} i\mu(\alpha_i)$

*Complexity* The number of operation performed in every node is proportional with the number of children of that node. Since the number of visited node is $O(n)$, this algorithm runs with $O(n)$ time complexity.

Figure 3 illustrates the computation of the maximum induced matching for the graph in Figure 2 using our algorithm. The set above every node represents the maximum induced matching of that node and the set under a node represents two nonadjacent vertices in this node.

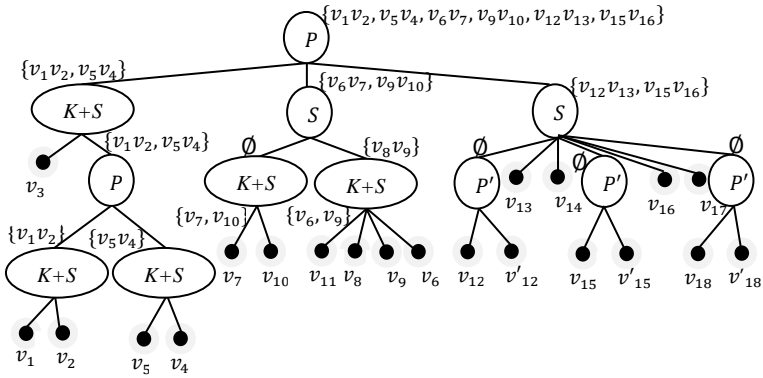

Figure. 3. The computation of the maximum induced matching for the graph in Figure 2.

## IV. Conclusion

The maximum induced matching algorithm is computed in $O(n)$ time, given a canonical decomposition tree of a bipartite $Star_{123}$-free graph. The canonical decomposition of a bipartite $Star_{123}$-free graph can be done in $O(n + m)$ time where $m$ is the number of edges [6]. Thus, the whole process is in $O(n + m)$ time.

## Acknowledgment

This research is funded by the Deanship of Research and Graduate Studies in Zarqa University /Jordan.